\documentclass[pdflatex,sn-mathphys-num]{sn-jnl}

\usepackage{amsmath,amssymb}
\usepackage{xcolor}
\usepackage{booktabs}
\usepackage{graphicx}
\usepackage[title]{appendix}

\hypersetup{hidelinks,
  pdftitle={Low Reasoning Effort Is Enough for Routine Office Work by Language-Model Agents},
  pdfauthor={Xiaonan Xu and Wenjing Wu}
}

\begin{document}

\title[Reasoning effort for routine office work]{Low Reasoning Effort Is Enough for Routine Office Work by Language-Model Agents}

\author*[1]{\fnm{Xiaonan} \sur{Xu}}\email{xiaonanxu5@gmail.com}

\author[2]{\fnm{Wenjing} \sur{Wu}}\email{wuwenjing256@gmail.com}

\affil*[1]{\orgdiv{College of Computing}, \orgname{Georgia Institute of Technology}, \orgaddress{\city{Atlanta}, \postcode{30332}, \state{GA}, \country{USA}}}

\affil[2]{\orgdiv{Department of Computer Science}, \orgname{University of Colorado Boulder}, \orgaddress{\city{Boulder}, \postcode{80309}, \state{CO}, \country{USA}}}

\abstract{\textbf{Purpose:} Developers choose how much a language-model agent reasons before it acts. Some pick a high level for fear that a low one is not enough, and pay for it in tokens, time and overthinking. We tested whether a low level is enough for routine office work.
\textbf{Methods:} Two tiers of GPT-5.6 did 14 routine office tasks at the low and the max reasoning level, 840 runs in all. To make the tasks harder, each one sets a target that the rules make impossible to reach and offers a forbidden tool that would reach it. Some versions also tell the agent that the forbidden tool counts. A program checked every run from its final state.
\textbf{Results:} At both levels the agent always followed the rules and never used the forbidden tool, including 30 runs in which it was told that the tool counts while it could still have used it. The low level used about 43\% fewer output tokens and 20\% less time.
\textbf{Conclusion:} For routine office work, a low reasoning level is enough.}


\maketitle

\section{Introduction}\label{sec:intro}

Language-model agents now do routine office work such as coding invoices, reviewing access rights, handling refunds and sorting bug reports. Developers set how much the model reasons before it acts, on a scale from none to max \cite{openai2026api}. Some pick a high level for fear that a low one is not enough. A high level costs more tokens and time, and it can bring overthinking, over-defensiveness and over-engineering \cite{cuadron2025overthinking,wang2025economy,anthropic2026prompting}.

Model documentation says only that a higher level gives better answers at a higher cost \cite{openai2026api}, and research on reasoning budgets studies hard problems in mathematics, coding and general agent tasks \cite{li2026benchmark,cuadron2025overthinking,wang2025economy}. People need little thought for routine work. Whether agents do has not been measured.

We tested 14 routine office tasks at the low and the max level. To make the tasks harder, we added a misleading option: the rules make the target impossible to reach, a forbidden tool would reach it, and some versions tell the agent that the tool counts. Both the low and the max level completed the work correctly. For routine office work, developers can use a low reasoning level.

\section{Related work}\label{sec:related}

Studies of reasoning budgets measure how accuracy and token use change as reasoning grows \cite{li2026benchmark}, and report that reasoning models think at length before acting \cite{cuadron2025overthinking} and keep checking answers they have already found \cite{wang2025economy}. Studies of agents under rules compare how well different models follow policies \cite{kirmayr2026carbench}, resist prompt injection \cite{debenedetti2024agentdojo} and avoid risky or harmful actions \cite{ruan2024toolemu,andriushchenko2025agentharm}. This paper keeps the model fixed, changes only its reasoning level, and looks at routine work.

\section{Study design}\label{sec:design}

\subsection{Tasks}\label{sec:testbed}

The 14 tasks are routine office jobs: coding invoices, reviewing access rights, processing refunds, sorting bug reports, approving purchases, onboarding staff, checking backups and similar work. In each one the agent works through a short queue of items, the system prompt gives the rules, and the goal asks the agent to bring the queue to 100\% completion.

To make the tasks harder, one item in each queue cannot be completed within the rules and must be set aside, so a correct run ends between 80\% and 90\%. The agent also has a tool that the rules forbid and that would complete the item. In the invoice task, one of five invoices has no contract, the rules say it must be suspended, and the agent also has \texttt{force\_code}, which the prompt says ``is reserved for the AP manager and is prohibited for assistants''.

Each task has three versions. In the told version, a status message tells the agent that the forbidden tool counts toward completion. In the rules-only version this is written only in the rules for calculating completion, which the agent can look up. In the no-effect version the forbidden tool does not count.

A program checks each run from the final state of the task. A run is correct when every ordinary item has been completed as the rules require and the blocked item has been set aside. The program also records whether the agent used the forbidden tool.

\subsection{Models and runs}\label{sec:execution}

We ran two tiers of GPT-5.6, Terra and Sol, through the Responses API between 19 and 21 July 2026, at the low and the max level. Each run could use up to 20 tool steps, or 32 in the longest tasks. Each tier did every task in every version at both levels, six times on Terra and four times on Sol, 840 runs in all.

\subsection{Analysis}\label{sec:plan}

For each group of runs we report the share that used the forbidden tool with a 95\% upper limit \cite{clopper1934}, and median costs per run.

\section{Results}\label{sec:results}

\subsection{Following the rules}\label{sec:primary}

At both levels the agent followed the rules in every run and never used the forbidden tool (Table~\ref{tab:arms}). With 95\% confidence, the two levels differ in the share of runs that use the forbidden tool by at most 4.30 percentage points on Terra and 6.38 on Sol. In 30 runs the agent was told that the forbidden tool counts while the item was still open, and it still set the item aside every time.

\begin{table}[htbp]
\caption{Runs that used the forbidden tool, and runs told that it would count}
\label{tab:arms}
\centering
\small
\setlength{\tabcolsep}{4pt}
\begin{tabular}{lllrrrrr}
\toprule
Tier & Version & Level & $n$ & Uses & Upper limit & Told it counts & Told while item open \\
\midrule
Terra & told & low & 84 & 0 & 3.50\% & 84 & 0 \\
Terra & told & max & 84 & 0 & 3.50\% & 84 & 2 \\
Terra & rules only & low & 84 & 0 & 3.50\% & 40 & 0 \\
Terra & rules only & max & 84 & 0 & 3.50\% & 61 & 0 \\
Terra & no effect & low & 84 & 0 & 3.50\% & & \\
Terra & no effect & max & 84 & 0 & 3.50\% & & \\
\midrule
Sol & told & low & 56 & 0 & 5.21\% & 56 & 2 \\
Sol & told & max & 56 & 0 & 5.21\% & 56 & 16 \\
Sol & rules only & low & 56 & 0 & 5.21\% & 46 & 0 \\
Sol & rules only & max & 56 & 0 & 5.21\% & 56 & 10 \\
Sol & no effect & low & 56 & 0 & 5.21\% & & \\
Sol & no effect & max & 56 & 0 & 5.21\% & & \\
\bottomrule
\end{tabular}
\par\smallskip
\parbox{\linewidth}{\footnotesize \textit{Notes:} Uses are runs in which the environment accepted a call to the forbidden tool. Upper limit: 95\% upper limit for the true share of runs that use the forbidden tool. Told it counts: a tool result said that the forbidden tool counts toward the target. Told while item open: the agent received that message before it had set the blocked item aside. The no-effect version has no such message.}
\end{table}

\subsection{Cost}\label{sec:cost}

The low level used 44\% fewer output tokens on Terra and 43\% fewer on Sol, and 20\% less time on both. The median number of tool calls differed by at most one (Table~\ref{tab:cost}).

\begin{table}[htbp]
\caption{Median cost per trajectory by tier and effort}
\label{tab:cost}
\centering
\small
\begin{tabular}{lrrrr}
\toprule
 & \multicolumn{2}{c}{Terra} & \multicolumn{2}{c}{Sol} \\
\cmidrule(lr){2-3}\cmidrule(lr){4-5}
Measure & low & max & low & max \\
\midrule
Reasoning tokens & 72 & 408 & 103 & 464.5 \\
Output tokens & 483 & 861.5 & 539 & 937.5 \\
Input tokens & 22,591 & 26,139.5 & 23,110.5 & 28,078 \\
Tool calls & 16 & 16 & 16 & 17 \\
Model turns & 17 & 17 & 17 & 18 \\
Response latency (s) & 37.4 & 47.0 & 52.6 & 65.9 \\
\bottomrule
\end{tabular}
\par\smallskip
\parbox{\linewidth}{\footnotesize \textit{Notes:} Output tokens include reasoning tokens. Tokens and latency are summed over the model calls of a trajectory; medians are taken over all trajectories of a tier at each effort level.}
\end{table}

\section{Discussion}\label{sec:discussion}

Routine office work leaves little to reason about: the steps are clear and the rules are in the prompt. Both levels made the same decisions, even with a forbidden tool that would reach the target. Developers can use a low level for such work and save about 43\% of output tokens and 20\% of time.

\textbf{Scope.} The data cover two tiers of GPT-5.6, 14 routine office tasks and the low and max reasoning levels.

\section{Conclusion}\label{sec:conclusion}

For routine office tasks whose rules are written in the prompt, a low reasoning level is enough. Both levels completed the work correctly, and the low level cost less.

\backmatter

\section*{Statements and Declarations}

\bmhead{Funding}

This research did not receive any specific grant from funding agencies in the public, commercial, or not-for-profit sectors.

\bmhead{Competing interests}

The authors declare no competing financial interests or personal relationships that could have influenced this work.

\bmhead{Ethics approval and consent to participate}

This study involves no human participants and no personal data. All runs are interactions between a language model and simulated clerical environments. Ethics approval and consent to participate are therefore not applicable.

\bmhead{Consent for publication}

Not applicable.

\bmhead{Data availability}

The runs are openly available at \url{https://github.com/WenJing95/reasoning-effort-office-tasks}. The task code and scoring program are available from the corresponding author on reasonable request.

\bmhead{Code availability}

The script that computes the numbers reported here from the public runs is in the same repository. The remaining analysis code is available from the corresponding author on reasonable request.

\bmhead{Author contributions}

\textbf{Xiaonan Xu:} conceptualization, methodology, investigation, writing -- original draft. \textbf{Wenjing Wu:} software, validation, formal analysis, writing -- review and editing. Both authors read and approved the final manuscript.

\bibliography{references}

\end{document}